\documentclass[aps,prd,twocolumn,nofootinbib,showpacs,preprintnumbers,showkeys]{revtex4}

\usepackage{epsfig}
\usepackage{amsmath}
\usepackage{amssymb}
\usepackage[dvipdfm]{hyperref}

\newcommand{\ppint}{\int_0^{2\pi}\hspace{-2.08em}\not\hspace{1.6em}}

\begin{document}

\preprint{ITEP--TH--22/11}

\title{The QCD scattering amplitude from area behaved Wilson loops}

\date{January 24, 2012}

\author{Y. Makeenko}

\affiliation{Institute of Theoretical and Experimental Physics,\\
B.~Cheremushkinskaya 25, 117218 Moscow, Russia}
\email{makeenko@itep.ru} 

\author{P. Olesen} 
\affiliation{The Niels Bohr International Academy,\\ The
Niels Bohr Institute,\\ Blegdamsvej 17, 2100 Copenhagen \O, Denmark}
\email{polesen@nbi.dk}

\begin{abstract}
We explicitly construct the dominant saddle-point trajectory in the 
sum-over-path
representation of meson scattering amplitudes in large $N$ QCD for area-behaved
Wilson loops and show that it dominates in the Regge regime. The graphic
representation of the leading trajectory is very similar to the diagrams
widely used to illustrate meson scattering.
\end{abstract}



\maketitle


In this Letter we study the large $N$ QCD scattering amplitudes of colorless
composite quark operators given as a path integral over Wilson loops 
\cite{w,mm}.
 We find the most important paths by means of a double saddle
point method. A geometric representation of these paths is very similar 
to the dual diagrams which for many years have been
used to illustrate meson scattering amplitudes.

Previously we have considered \cite{us} 
the large $N$ (or quenched) QCD meson scattering 
amplitude, using the connection between this amplitude and the Wilson loop
in the case where the loop is well represented by the area behavior.
This QCD scattering amplitude then turns out to be 
the Regge limit of the usual Veneziano amplitude 
in the cases when the number of external particles is large, or
when the quark mass is small relative to the QCD-$\Lambda$. As discussed in
\cite{us} this is a prediction for near forward scattering at high energies, 
that is associated with large distances in the relevant functional integral.
The goal of this Letter is to demonstrate this by explicitly constructing
the saddle point trajectory. 

The relation between the off shell $M$ meson scattering amplitude $G$ 
and the Wilson loop occurs as a functional integral~\cite{us},
\begin{eqnarray}
G(p_1,...,p_M)&\propto &\prod_{i=1}^{M-1}\int_0^{\phi_{i+1}} d\phi_i\int
{\cal D}x(\phi)\, e^{i\int d\phi\, p(\phi)\,\dot{x}(\phi)~} \nonumber \\
&&\hspace*{.4cm} \times W(x(\phi)), \quad x(0)= x(2\pi),
\label{1}
\end{eqnarray}
in the limit of large $M$ or small quark masses, when the spin dependence 
factorizes.
In this expression $W(x(\phi))$ is the Wilson loop defined on the 
boundary curve $x(\phi)$, 
and the function 
$p(\phi)$ is piecewise constant, defined by
\begin{equation}
p(\phi)=q_i~~{\rm for}~~\phi_i <\phi <\phi_{i+1},
\label{2}
\end{equation}
where the momenta $q_i$ are related to the external momenta  $p_i$'s  by
\begin{equation}
p_i=q_{i-1}-q_i,
\end{equation}
thus ensuring conservation of the energy-momentum. In particular we note that
Eq.~(\ref{2}) implies the very useful result
\begin{equation}
\dot{p}(\phi)=-\sum_i p_i\,\delta(\phi-\phi_i).
\label{4}
\end{equation}
Also, in Eq.~(\ref{1}) the Wilson loop is required to be bounded by the closed
curve $x^\mu (\phi)$.

In the basic relation (\ref{1}) we now want to insert the area behavior
$W\sim\;$exp(-area), 
where the area is measured in terms of the string tension.
If we insert the area as the usual square root of the determinant of
the extrinsic metric and minimize this, the functional {\it path}\/ integral 
in (\ref{1})
becomes impossible to perform because it depends on surface variables instead 
of the paths in (\ref{1}) and also because 
of the highly non-linear character of the
integration. We therefore suggested \cite{us} to use Douglas' expression
for the minimal area \cite{douglas},
\begin{eqnarray}
A[x]&=&\frac{-1}{4\pi}\int_0^{2\pi}d\theta_1\int_0^{2\pi}d\theta_2\,
\dot{x}(\theta_1)\,\dot{x}(\theta_2)\,\nonumber \\*
&& \hspace*{5mm} \times \log [1-\cos(\phi(\theta_1)-
\phi(\theta_2))].
\label{5}
\end{eqnarray}
This integral should be minimized with respect to the reparametrization 
function $\theta (\phi)$, subject to the requirement $\theta(\phi_i)=\phi_i$.
We should therefore look for a saddle point with respect to the
reparametrization.

The expression (\ref{5}) is known from the derivation of
the string disk amplitude in the Polyakov formulation,
where it appeared as the boundary action after integration over
fluctuations of an open string with the Dirichlet boundary conditions
at the ends. It is less known that the crucial role is then played by
the path integral over the boundary metric (the same
as the path integral over the reparametrizations), which
is dominated for large loops by the saddle point in hands,
reproducing the minimal area. 
However, for small loops $W(x(\phi))$ in Eq.~(\ref{1}) is described by
QCD perturbation theory, rather than by the string disk amplitude, 
thus leading us to the understanding~\cite{Ole85} of QCD string as 
an effective rather than fundamental string. It makes sense only for long
strings and we shall soon justify by an explicit calculation that the 
double saddle point we consider is indeed associated 
with large loops or long QCD string.

Douglas' integral (\ref{5}) is apparently well suited for insertion in the 
{\it path}\/ integral (\ref{1}), where it leads to an apparent Gaussian 
integration. However, this point of view is an illusion -- the non-linearity is
still with us, because we have to minimize the integral (\ref{5}) with respect
to reparametrizations $\theta(\phi)$ or the inverse $\phi (\theta)$. As
shown by Douglas, this leads to the requirement
\begin{equation}
\dot{x} (\theta_1) 
\ppint  d\theta_2 \,\dot{x}(\theta_2)\,\cot
\left(\frac{\phi(\theta_1)-\phi (\theta_2)}{2}\right)=0.
\label{6}
\end{equation}
This condition is highly non-linear!

We now proceed to tackle this problem by a two step mechanism: first we 
find the saddle point in the $x-$integral for any reparametrization (justified
because we seek a large minimal area), and then 
we require that for this saddle point the reparametrization must satisfy  
the condition (\ref{6}). Thus we search for a double saddle point.

From Eq.~(\ref{1}) the saddle point condition  for large area or, 
alternatively, large string tension is
\begin{equation}
ip^\mu(\theta)+\frac{\sigma}{2\pi}\int_0^{2\pi}d\theta'\,\dot{x}^\mu(\theta')\,\log
[1-\cos(\phi(\theta)-\phi(\theta'))]=0
\label{7}
\end{equation}
for $W\sim \exp(-\sigma A)$, where $\sigma$ is the string tension.
This equation can easily be solved by noting that the log is a Green function
with a simple inverse,
\begin{subequations}
\begin{eqnarray}
G(\phi)&\equiv &\frac{-1}{2\pi}~\log (2-2\cos\phi),
\label{8a}\\
G^{-1}(\phi)&=&
-\ddot{G}(\phi)=\frac{-1}{2\pi (1-\cos\phi)}.
\label{8b}
\end{eqnarray}
\label{8}
\end{subequations}
Inverting (\ref{7}) we therefore get
\begin{equation}
x^\mu (\theta)=\frac{i}{2\pi \sigma}\int_0^{2\pi}d\theta'\,\dot{p}^\mu 
(\theta')\,\log [1-\cos(\phi(\theta)-\phi(\theta'))].
\label{9}
\end{equation}
Note that this saddle point $x$ is imaginary, as is to be expected from the
occurrence of $ipx$ in (\ref{1}). Next we can use Eq.~(\ref{4}) to obtain
\begin{equation}
x^\mu (\theta (\phi))=\frac{-i}{2\pi\sigma}~\sum_j~p_j^\mu\log[1-\cos (\phi
-\phi_j)].
\label{10}
\end{equation}
We note the subtle point that in principle the right-hand side depends on
the reparametrization, but the left-hand side is independent! Thus, the saddle 
point does not care about $\theta (\phi)$, i.e. the reparametrization can be 
any function. 

We now have to show that the saddle point (\ref{10}) actually satisfies the 
non-linear Douglas condition (\ref{6}). From (\ref{10}) we have
\begin{equation}
\dot{x}^\mu(\theta)=\frac{-i}{2\pi\sigma}\dot{\phi}(\theta)\sum_j p_j^\mu\,\cot
\left(\frac{\phi(\theta)-\phi(\theta_j)}{2}\right).
\end{equation}
Inserting this on the left-hand side of Eq.~(\ref{6}), we obtain
\begin{equation}
\dot{x}(\theta)\sum_jp_j 
\ppint d\phi'\,\cot\left(\frac{\phi'-\phi_j}
{2}\right)\cot\left(\frac{\phi'-\phi(\theta)}{2}\right)
\stackrel{{\large ?}}=0. 
\label{12}
\end{equation}

The left hand-side can be easily evaluated if we perform a partial integration
and use Eqs.~(\ref{8a}) and (\ref{8b}), 
\begin{eqnarray}
\lefteqn{\hbox{left-hand side of (12)}}\nonumber \\*
&\propto &\dot{x}(\theta)\sum_j~p_j
\ppint d \phi'\,
G(\phi'-\phi_j)~G^{-1}(\phi'-\phi(\theta))\nonumber \\
&=&\dot{x}(\theta)\sum_j p_j\,\delta (\phi (\theta)-\phi_j).
\label{c1}
\end{eqnarray}
In order to satisfy the Douglas condition (\ref{6}) we therefore need to have
\begin{equation}
\sum_i p_ip_j \cot\left(\frac{\phi_i-\phi_j}{2}\right)=0
\label{c2}
\end{equation}
for each value of $j$. Among these equations only $M\!-\!3$ are
independent because of the projective symmetry,
thus fixing $M\!-\!3$
integration variables $\phi_i$'s as a consequence of the fact that we are 
looking for a double saddle point. The remaining three variables can be fixed
arbitrarily.

Instead of the angle $\phi$ ranging between 0 and $2\pi$ it is often convenient
to use 
\begin{equation}
r=-\cot (\phi/2),
\end{equation}
which ranges from $-\infty$ to $+\infty$. The saddle point solution (\ref{10})
then becomes
\begin{equation}
x^\mu (u(r))=\frac{-i}{\pi\sigma}\sum_jp_j^\mu\,\log |r-r_j|
\label{16}.
\end{equation}
Here $u(r)$ is any reparametrization of the boundary curve.

Since the trajectory (\ref{16}) is the saddle point of the amplitude (\ref{1}),
it can be approximated by the value of the integrand at this saddle point.
We thus obtain~\cite{us}
\begin{eqnarray}
\lefteqn{e^{-\sigma A[x]+i\int p(r)\dot{x}(r)dr}|_{\rm saddle~point~value}}  \nonumber\\
&&=e^{(1/2\pi\sigma)\,\sum_{i,j} \,p_ip_j\,\log|r_i-r_j|} ~~~~~
\label{17}
\end{eqnarray}
which coincides with the saddle point approximation to
the integral over 
the Koba-Nielsen variables in the $M$ point Veneziano amplitude.

This stems from the fact that the Douglas procedure leads to a 
determination of the parameters $r_i$'s in a similar way how we
have seen this in Eq.~(\ref{c2})
for the corresponding parameters $\phi_i$'s. 
Proceeding as before we 
obtain~\cite{Mak11}
\begin{equation}
\sum_i\frac{p_ip_j}{r_i-r_j}=0.
\label{c3}
\end{equation}
The meaning of this condition is easy to understand:
on the right-hand side of Eq.~(\ref{17}) the condition 
(\ref{c3}) corresponds to the 
extremum, generally including the Regge limit, and this extremum is thus 
exactly the minimal surface entering the Wilson loop in QCD! 

\begin{figure}
\vspace*{1.mm}
\includegraphics[width=5.9cm]{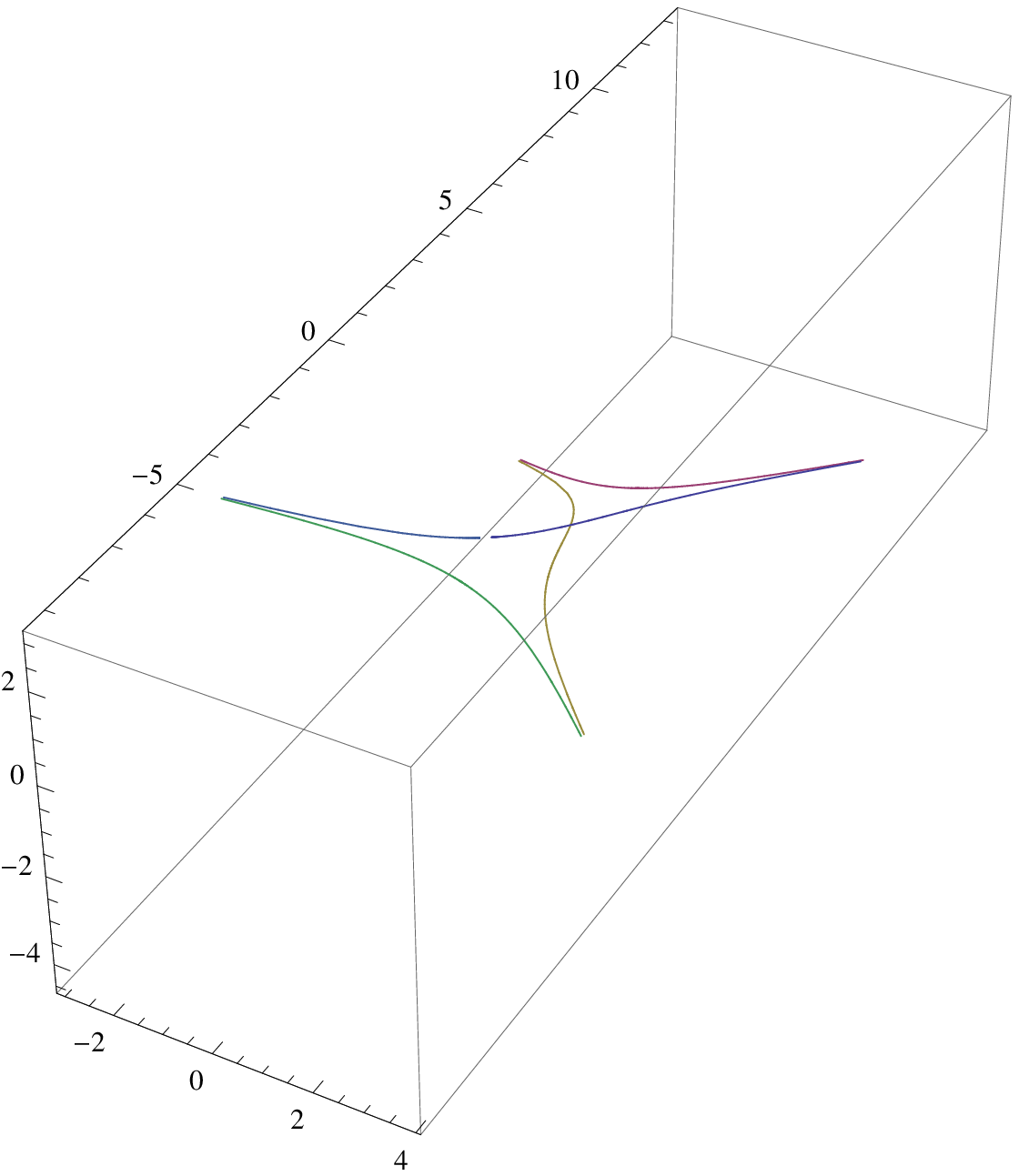} \\ 
\includegraphics[width=5.0cm]{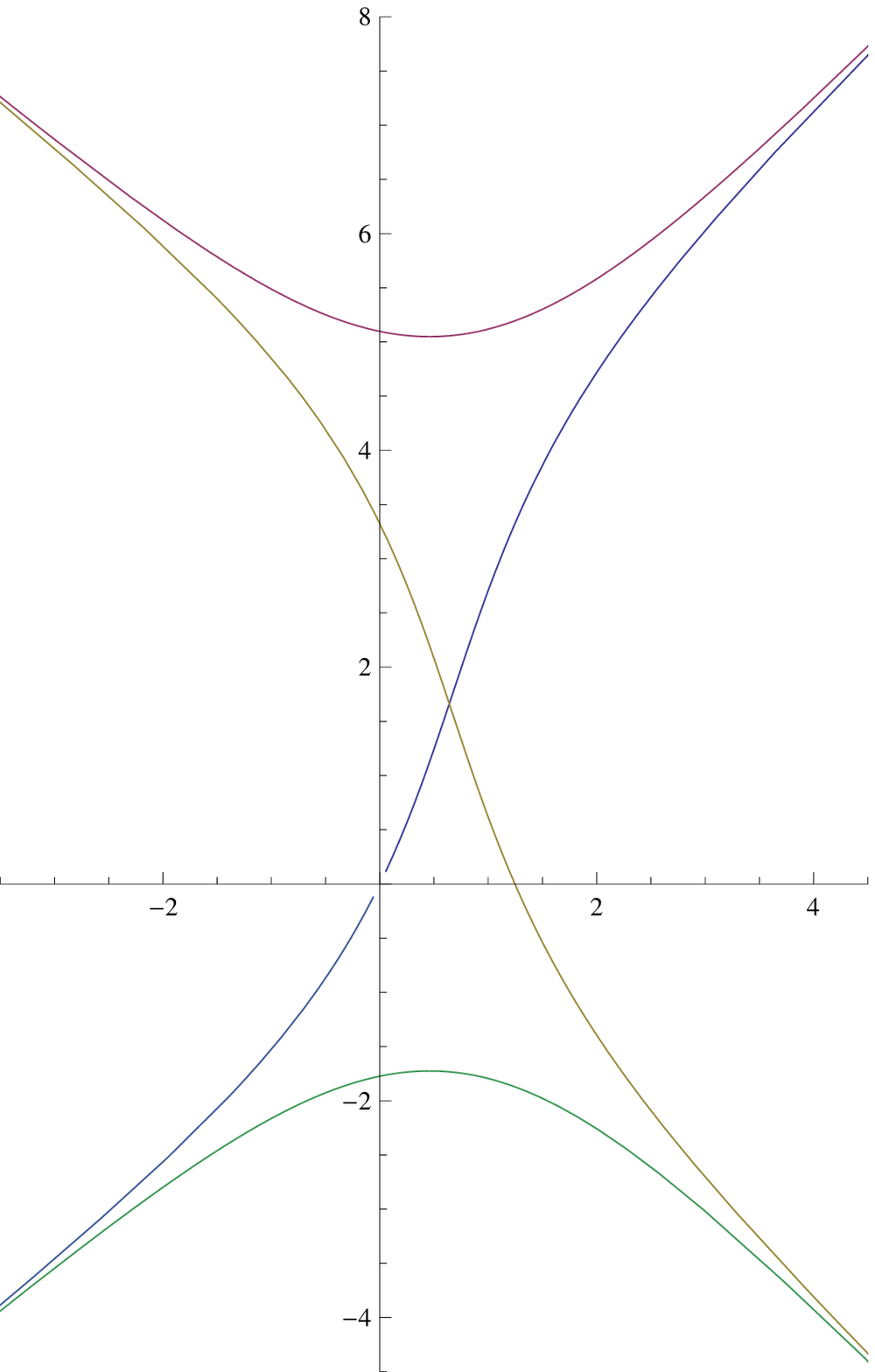} \\ 
\includegraphics[width=5.8cm]{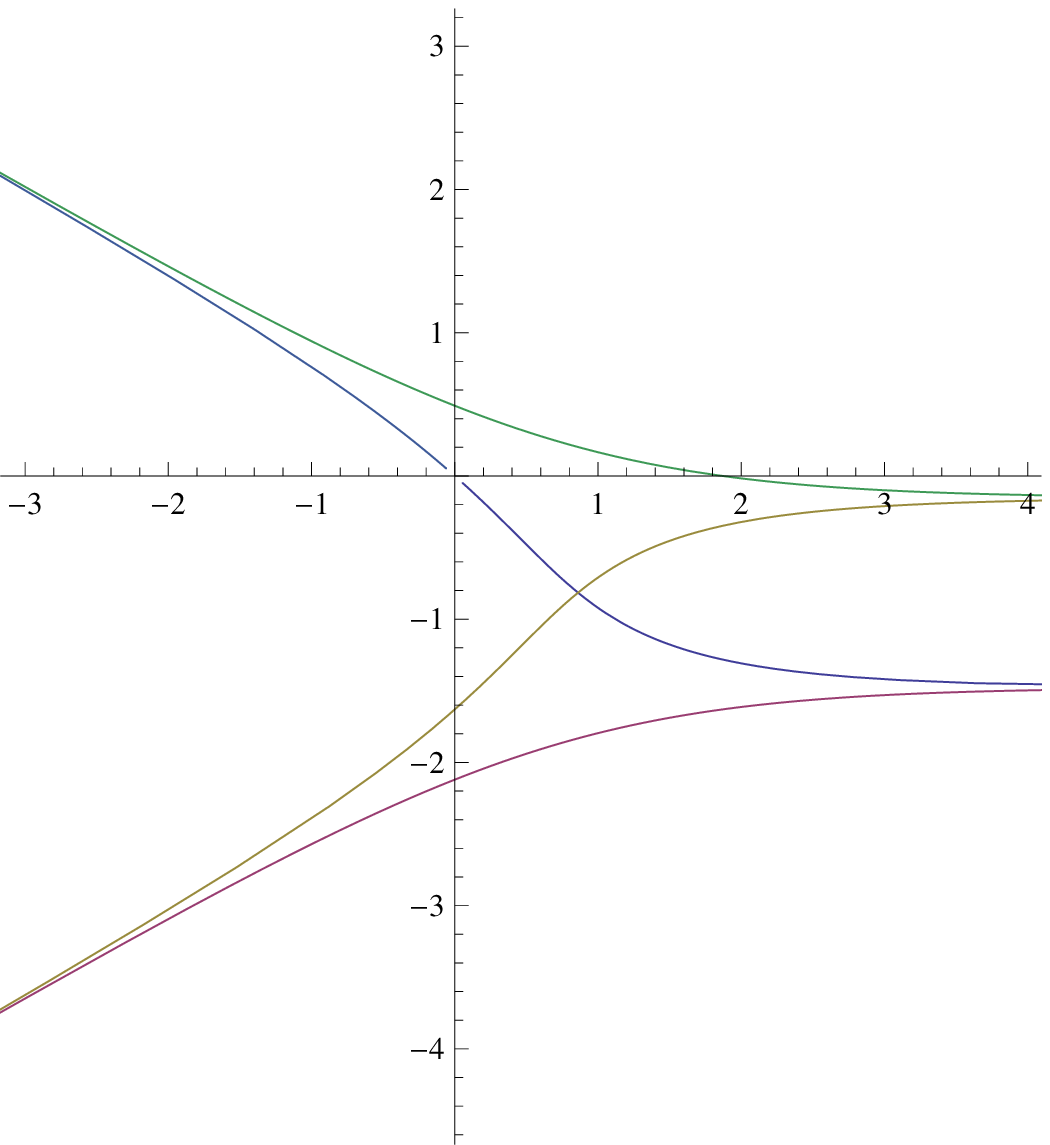} \\
\includegraphics[height=2.66cm,width=6cm]{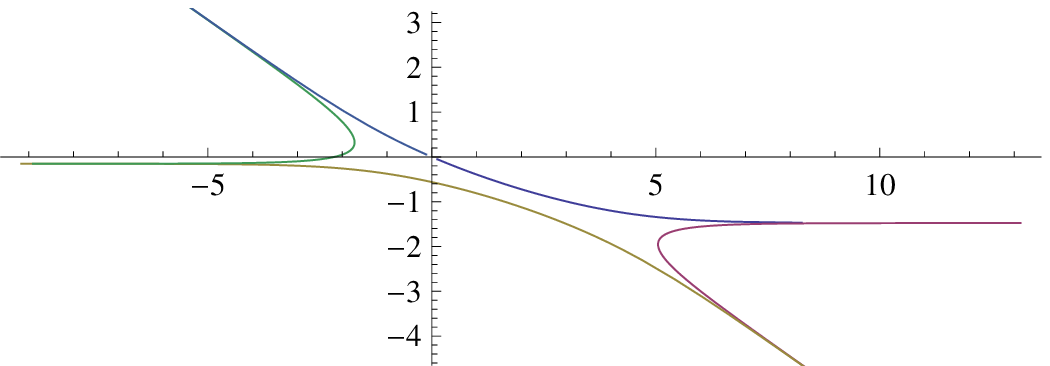}
 \caption[]{Saddle-point trajectory (\ref{16}) for $t/s=0.1$:
3D view and its projections in 0,1; 0,2 and 1,2 planes, respectively.}   
\label{fi:point1}
\end{figure}

It is instructive to see how the curve~(\ref{16}) looks like for the
four point function in
the center of mass frame. Because we deal with the case where both
Mandelstam's variables $s$ and $t$ are negative, this is associated with
a scattering process in $u$-channel. 
In view of the projective symmetry we can set three out of four $r_j$'s
equal to certain typical values, keeping in mind that one
such contour belongs, in fact, to a whole family of contours,
which are equally important. The projective invariant ratio
$(r_2-r_1)(r_4-r_3)/(r_3-r_1)(r_4-r_2)$ is then fixed by Eq.~(\ref{c3})
to be $t/(t+s)$. We plot the corresponding contour
in Fig.~\ref{fi:point1} for $s/t=10$ in the units of $i\sqrt{u}/2\pi\sigma$.
The contours are quite normal for not too small $t/s$ 
and look pretty much similar 
to dual diagrams used to illustrate the scattering of mesons.
For $r\to r_j$ we have
\begin{equation}
x^\mu(u(r))\to \frac{-i}{\pi\sigma}p^\mu_j \times \,\hbox{regularizing factor}. 
\end{equation} 
Actually, Fig.~\ref{fi:point1} is obtained without any regularization,
which is implemented by {\sc Mathematica} itself.
The transverse size of the contour shrinks when $t/s\to0$
in spite of the fact that the minimal area
\begin{equation}
\sigma S_{\rm min}(C)=\alpha' t \log \frac st, \qquad 
\alpha'=\frac{1}{2\pi\sigma}, 
\label{Amin}
\end{equation}
can be large even for small $t$.
For the self-consistency of our approach this transverse size
should be larger than the confinement scale 
$1/\Lambda_{\rm QCD}\sim 1/\sqrt{\sigma}$.

A very interesting question is as to how we can make the saddle
point trajectory (16) to be real. This happens in the physical domain
of $s$-channel, when $s>0$, $t<0$ and $u<0$, so that $-i \sqrt{u}=\sqrt{s}$.
We may think about this as an analytic continuation in $s$ from negative
to positive values, keeping $|s|\gg -t$ all the way. The area of such 
a surface will be a complex number because of Minkowski-space kinematics but 
with a negative real part, as is seen from Eq.~(\ref{Amin}), what is
essential in our consideration for it to dominate in the path integral.

There is a number of well-known reasons why the amplitude~(\ref{17}) cannot be 
right in all kinematical regions of QCD. For example, for large
transverse momenta the decay should be power like, in contrast to
the dual models, and furthermore, QCD has of course no tachyon. 
In this connection it should be remembered that the area behaved 
Wilson loop is quite small, so it may only dominate in restricted kinematical 
regions. For large transverse momenta the amplitude corresponding to
Eq.~(\ref{17}) decays exponentially,
so the power behaved perturbative behavior of $W$ is much more important for 
large transverse momenta. We therefore expect the area behavior to dominate 
only at 
large energies in the near forward direction. Since the tachyon is a low 
energy phenomenon~\cite{Ole85} it is therefore not relevant in our work. 

From Eq.~(\ref{16}) we can obtain the equation for the minimal surface because 
$x$ has to be harmonic,
\begin{equation}
x^\mu (r,y)=\frac{-i}{\pi\sigma}\sum_j p^\mu_j \log |z-r_j|,\qquad z=r+iy,
~~~y\geq 0.
\label{log}
\end{equation}
This expression is the imaginary part of the analytic function
\begin{equation}
\frac{-i}{\pi\sigma}\sum_j p^\mu_j \log (z-r_j),
\end{equation}
whose real part 
\begin{equation}
\frac{1}{\pi\sigma}\sum_j p^\mu_j \arctan\frac{r-r_j}{y}
\label{arctan}
\end{equation}
is also harmonic and describes~\cite{Mak11} the minimal surface 
bounded by a step function
\begin{equation}
\frac{1}{\sigma}\sum_j p^\mu_j \Theta (r-r_j)\Theta (r_{j+1}-r)
\end{equation}
with $\Theta$ being the Heaviside step function.
It is of course related to the imaginary part by the 
Cauchy-Riemann equations, which 
in turn are related to a T-duality map \cite{pawel}.

The induced metrics of the surfaces (\ref{log}) and (\ref{arctan})
differ only by the signs.
Moreover, if we were go beyond the saddle point approximation,
their transverse quantum fluctuations can be similarly taken
into account by a conformal mapping of the upper half-plane
onto a rectangle, associated with the worldsheet parametrization.
We may expect therefore that we shall obtain for (\ref{log}), 
like in \cite{Mak11} for (\ref{arctan}), the linear Regge trajectory
\begin{equation}
\alpha(t)=\frac{d-2}{24} +\alpha' t
\label{linear}
\end{equation}
in the semiclassical approximation for arbitrary dimension $d\geq2$. 

The expression (\ref{17}) leads to straight line Regge trajectories,
when the values of $r_i$'s obey Eq.~(\ref{c3}). 
The prediction is thus that
this should occur at high energies and moderate transverse momenta, in 
accordance with experimental results
for SU(3). Also our results are valid in other dimensions than four, for
example in two dimensions, where 't Hooft \cite{th} found that the mass
spectrum for large masses is a linear function of an integer quantum number. 
For smaller masses there is a deviation from the straight line. In
two dimensions the Wilson loop is area behaved everywhere, but our calculation
only includes the saddle point. For smaller areas there would be corrections
to what we found.

The fact that the minimal surface (\ref{log}) is associated with 
the Regge limit of the Koba--Nielsen
amplitudes is known since early days of string theory \cite{HSV70}, though
the reparametrization of the boundary was not taken into account that time.
To get it within our approach, it was crucial that the path
integral (\ref{1}) is dominated by the saddle point trajectory~(\ref{16})
whose shape is not sensitive to the choice of the reparametrizing 
function $u(r)$. It is also worth mentioning another minimal surface 
of the form of a helicoid
that was associated in Ref.~\cite{JP00} with the reggeization 
of quark-antiquark 
within the AdS/CFT correspondence in a confining background.
It will be interesting to understand how these two are related.
However, we emphasize once again that our approach is based solely on
the representation of the Wilson loop via the boundary functional (\ref{5})
and does not use explicitly a surface representation.

As a conclusion we see that the large $N$ meson scattering amplitude expressed
in terms of a path integral is dominated by a double saddle point for
large energies and moderate transverse momenta. The graphic representation of
the leading trajectories is similar to the dual diagrams used to illustrate
meson scattering.

\end{document}